\documentclass{revtex4}
\usepackage{latexsym}
\usepackage{amsfonts}
\usepackage[latin1]{inputenc}
\usepackage[caption=false]{subfig}
\usepackage{graphicx}
\usepackage{mathrsfs}
\usepackage{amssymb}
\usepackage{hyperref}
\usepackage{amsmath}
\usepackage{verbatim}
\usepackage{multirow}
\usepackage{color}
\usepackage{epsfig}
\usepackage{amscd}
\usepackage{bm}
\usepackage{fancyhdr}
\usepackage{color}

\begin{document}

\title{Review of Higgs-to-light-Higgs searches at the LHC}

%

\author{D. Barducci$^{a}$, R. Aggleton$^{b,c,d}$, N-E Bomark$^{e}$, S. Moretti$^{b,c}$ and C. Shepherd-Themistocleous$^{e}$} 
\affiliation{
(a) SISSA/ISAS and INFN, 34136, Trieste, Italy \\
(b) Particle Physics Department, Rutherford Appleton Laboratory, Chilton, Didcot, Oxon OX11 0QX, UK   \\    
(c) School of Physics \& Astronomy, University of Southampton, Highfield, Southampton SO17 1BJ, UK  \\ 
(d) H. H. Wills Physics Laboratory, Bristol University, Bristol BS8 1TH, UK \\  
(e) Department of Natural Sciences, University of Agder, Postboks 422, 4604 Kristiansand, Norway
}

\begin{abstract}
We review the most relevant LHC searches at $\sqrt{s}=8$ TeV looking for low mass bosons arising from exotic decay of the Standard Model Higgs and highlighting their impact on both supersymmetric and not supersymmetric Beyond the Standard Model scenarios.
\end{abstract}

\maketitle

\thispagestyle{fancy}

\section{Introduction}
While the profile of the 125 GeV boson recently discovered at the LHC~\cite{Aad:2012tfa,Chatrchyan:2012xdj} is largely consistent with the Standard Model (SM) predictions, the possibility that this state belongs to a Beyond the Standard Model (BSM) scenario is one of the most pressing questions for particle physics, given the need of New Physics manifestations motivated from both the theoretical (hierarchy problem, absence of coupling unification...) and experimental  (neutrino masses, dark matter...) point of view.

Various BSM scenarios, whether or not constructed following the Naturalness paradigm, predicts the existence of an extended scalar sector, thus giving rise to additional physical Higgs states other than just the SM one. All these emerging Higgs bosons can, and in general do, interact with each other affording one the possibility of investigating the sector responsible for electro-weak symmetry breaking  in multiple ways.

In this proceeding, based on~\cite{Aggleton:2016tdd}, we review the experimental status of several BSM scenarios predicting exotic decays of the SM Higgs boson in a pair of light scalars (hereafter dubbed $a_1$ or $A$), with their subsequent decay into four SM states via the $h^{\rm SM}\to A A \to 4\times {\rm SM}$ chain.
We will adopt public data obtained during the 8 TeV run of the LHC from both the ATLAS and CMS collaborations, covering several signatures of such a pair of extra Higgs states, including decays into pairs of muons, taus and bottom quarks and showing their impact on the most popular BSM  models with an extended Higgs sector in which such a light object is realised.

\section{Experimental analyses}
\label{Sec:Exp}

There are several experimental analyses targeting light scalar bosons arising from the decay of the SM Higgs. One characteristic of these search channels is that, given the condition $m_A < m_{h^{\rm SM}}/2$, the $A$ bosons tends to be boosted, with a transverse momentum of order $p_T^A\sim m_{h^{\rm SM}}/2$, thus producing collimated decay products, with  $\Delta R\sim 2 m_A / p_T^A \sim 4 m_A / m_{h^{\rm SM}}$, therefore making these analyses challenging from an experimental point of view.
We here provide an overview of the most relevant search channels, categorised by their final state, and express the experimental limits in the equivalent $4\tau$ cross section by using the relations

\begin{equation}
\begin{split}
& \frac{BR(A\to\tau\tau)}{BR(A\to \mu\mu)}=\frac{m^2_\tau \beta(m_\tau, m_A)}{m^2_\mu \beta(m_\mu, m_A)} \\
& \frac{BR(A\to\tau\tau)}{BR(A\to b\bar b)}=\frac{m^2_\tau \beta(m_\tau, m_A)}{3 m^2_b \beta(m_b, m_A)(1+\Delta_{\rm rad})}
\end{split}
\label{eq:rescaling}
\end{equation}
with 

\begin{equation}
\begin{split}
\beta(m_X,m_A)=\sqrt{1-\left(\frac{2 m_X}{m_A}\right)^2}
\end{split}
\end{equation}
and $\Delta_{\rm rad}$ the radiative corrections to the $A\to b\bar b$ decay for which we refer to~\cite{Aggleton:2016tdd} for the complete expressions. Note that 
Eq.~\eqref{eq:rescaling} is valid for models in which all leptons and down-type quarks couple to the same Higgs doublet. We will therefore restrict our investigation to these classes of BSM scenarios.

\subsection{4$\tau$}

In the mass region $m_A \in [2 m_\tau - 2 m_b]$, the decay $A\to \tau\tau$ is typically expected to dominate and pair of ditaus are therefore a natural search channel, despite the
difficulty in reconstructing these states in a boosted regime.
The CMS collaboration has published two analyses that search for 4$\tau$ final states~\cite{Khachatryan:2015nba,CMS:2015iga} arising from the decays of a pair of low mass bosons, which differ for the strategies adopted to identify the boosted tau pairs and for the targeted production channels (either gluon fusion or gluon fusion combined with Higgs production in association with a $W$ boson). The limits set by these analysis on  the $4\tau$ cross section ranges from 10 pb for $m_A=5$ GeV to 3.5 pb for $m_A=$ 11 GeV. 

\subsection{2$\tau$2$\mu$}

This final state is a compromise between the dominant but less clean $4\tau$ final state, and the much cleaner but rarer $4\mu$ final state. Both ATLAS~\cite{CMS:2016cqw} and CMS~\cite{Aad:2015oqa} have searched for this channels looking at complementary mass region, $m_A \in [3.5 - 5]$ GeV for the former, and $m_A \in [20 - 62.5]$ GeV for latter. Differently from the ATLAS analysis, the CMS one, targeting heavier objects, does not require boosted techniques, and standard $\tau$ reconstruction algorithm and isolation requirements can be used. The obtained limits on  $\sigma\times {\rm BR}$ for the equivalent $4\tau$ final state ranges from 1 pb to 30 pb for ATLAS  and from 2 pb to 0.8 pb for the CMS case, where the quoted numbers refer to the extreme values considered for the $A$ mass. 

\subsection{4$\mu$}

In the region $m_A< 2 m_\tau$ a great increase in the $A\to \mu\mu$ decay rate is expected. While generally not as large as hadronic $A$ decays, the $4\mu$ topology takes advantages the very clean final state with reduced systematic uncertainties. CMS has searched for a $4\mu$ final state in the mass range $m_A \in [0.25 - 3.55]$ GeV setting a limit on the equivalent $4\tau$ cross section of $\sim$ 0.7 - 0.9 fb~\cite{Khachatryan:2015wka}.

\subsection{2$b$2$\mu$}

Above the $A\to b \bar b$ threshold, a 4$b$ final state generally becomes the dominant decay mode. However, this final state would have to overcome significant QCD backgrounds, so that requiring one $A$ to decay to a $\mu\mu$ pair will allow one to use the dimuon invariant mass as a discrimination between the signal and the background. CMS has searched for this decay topology in the mass range $m_A \in [25 - 65]$ GeV, thus exploiting standard particle reconstruction algorithms due to the non boosted regime of the decay products.
The limit on the equivalent $4\tau$ cross section ranges from 40 to 100 pb~\cite{CMS:2016cel}.

\section{Impact on multi Higgs models}
\label{Sec:Exp}

We now show how these constraints affect the parameter space of two specific multi Higgs models, namely the Two Higgs Doublet Model (2HDM) and the Next to Minimal Supersymmetric Standard Model (NMSSM). For the latter we consider both the standard  $Z_3$ invariant realisation as well as a non-$Z_3$ invariant variation, called new Minimal Supersymmetric Standard Model  (nMSSM). We refer the reader to~\cite{Branco:2011iw} and \cite{Ellwanger:2009dp,Barducci:2015zna} respectively for revies of the 2HDM and NMSSM/nMSSM scenario.
We have scanned the NMSSM parameter space by using the {\tt NMSSMTools}~\cite{Ellwanger:2005dv} code and imposing Higgs measurements constraints both via the {\tt HiggsSignals}~\cite{Bechtle:2013xfa} and {\tt HiggsBounds}~\cite{Bechtle:2011sb} packages, with both weak scale input parameters and universal parameters defined at the GUT scale, while we have exploited the {\tt 2HDMC}~\cite{Eriksson:2009ws} routine for scans in the 2HDM parameter space.

\begin{figure}[htb]
 \includegraphics[width=0.48\textwidth]{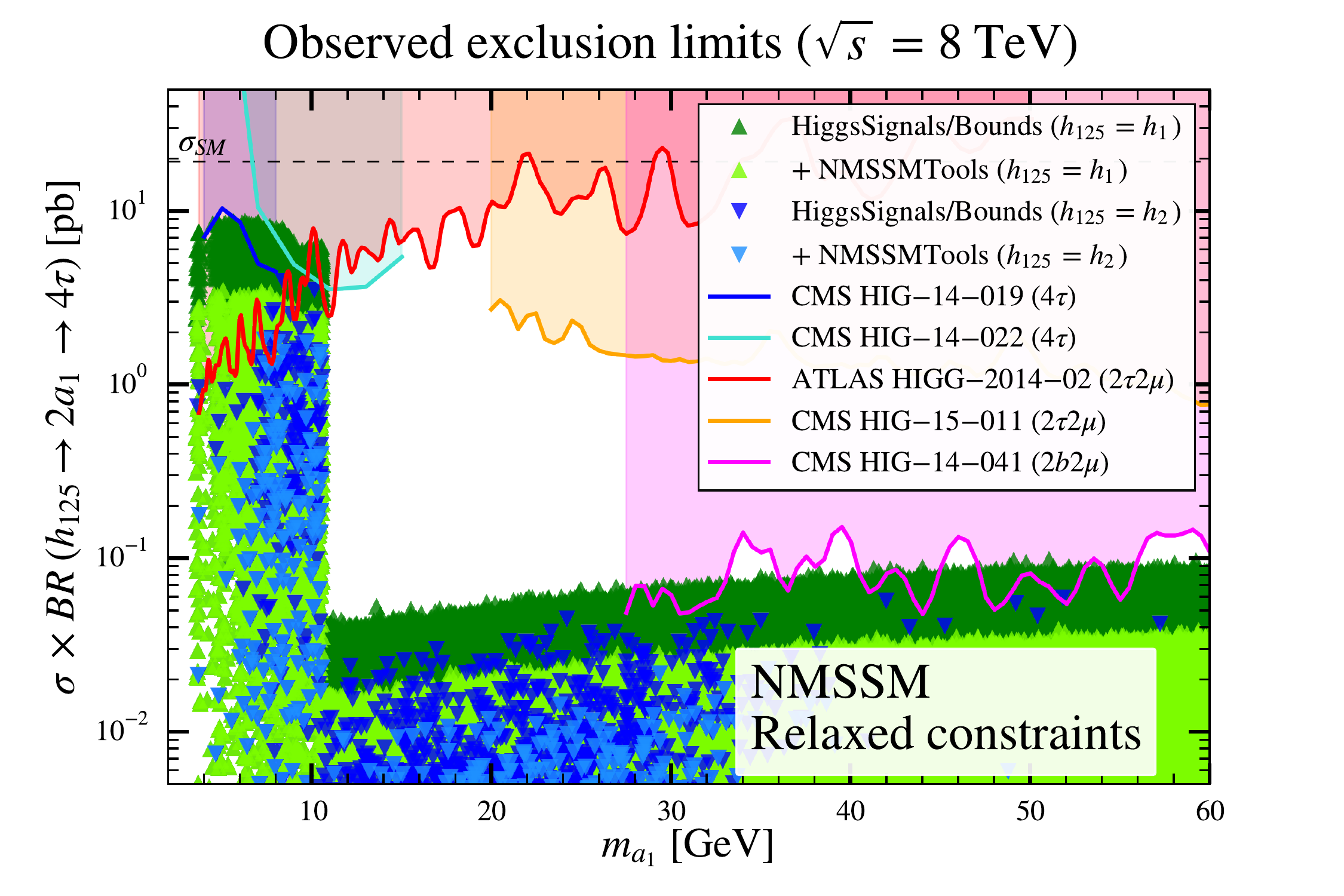}{}\hfill
  \includegraphics[width=0.48\textwidth]{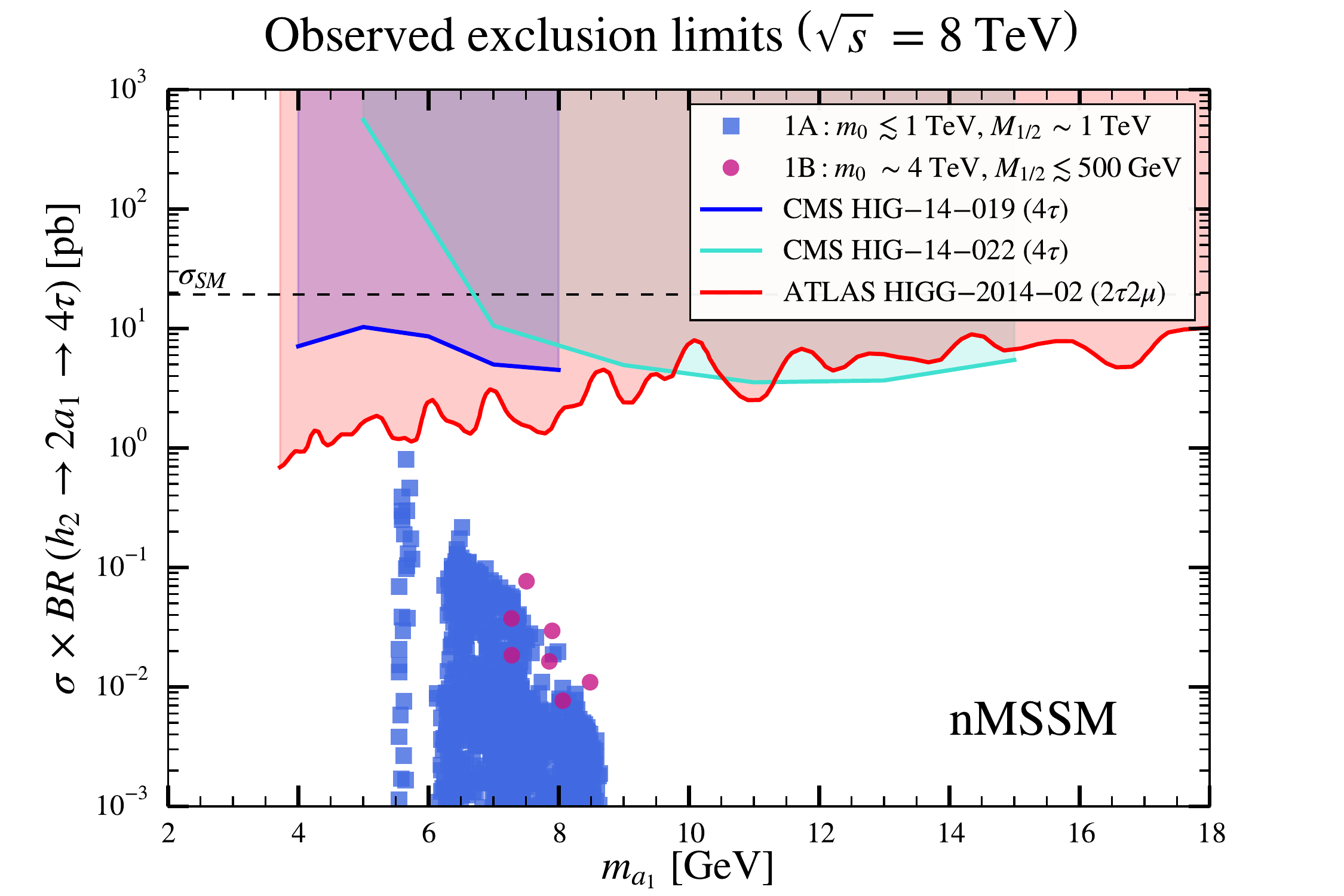}{}l
 \caption{Plots of $\sigma \times BR(h^{\rm SM} \to 2 A \to 4 \tau)$ versus $m_{a_1}$ for various Higgs assignments in the NMSSM. 
The left panel refers to the $Z_3$ invariant NMSSM with weak scale input parameters, while the right plot refers to the nMSSM with GUT scale input parameters.  We refer to~\cite{Aggleton:2016tdd} for details of the scan constraints indicated in the plot.}
 \label{fig:nmssm}
 \end{figure}

\begin{figure}[htb]
 \includegraphics[width=0.48\textwidth]{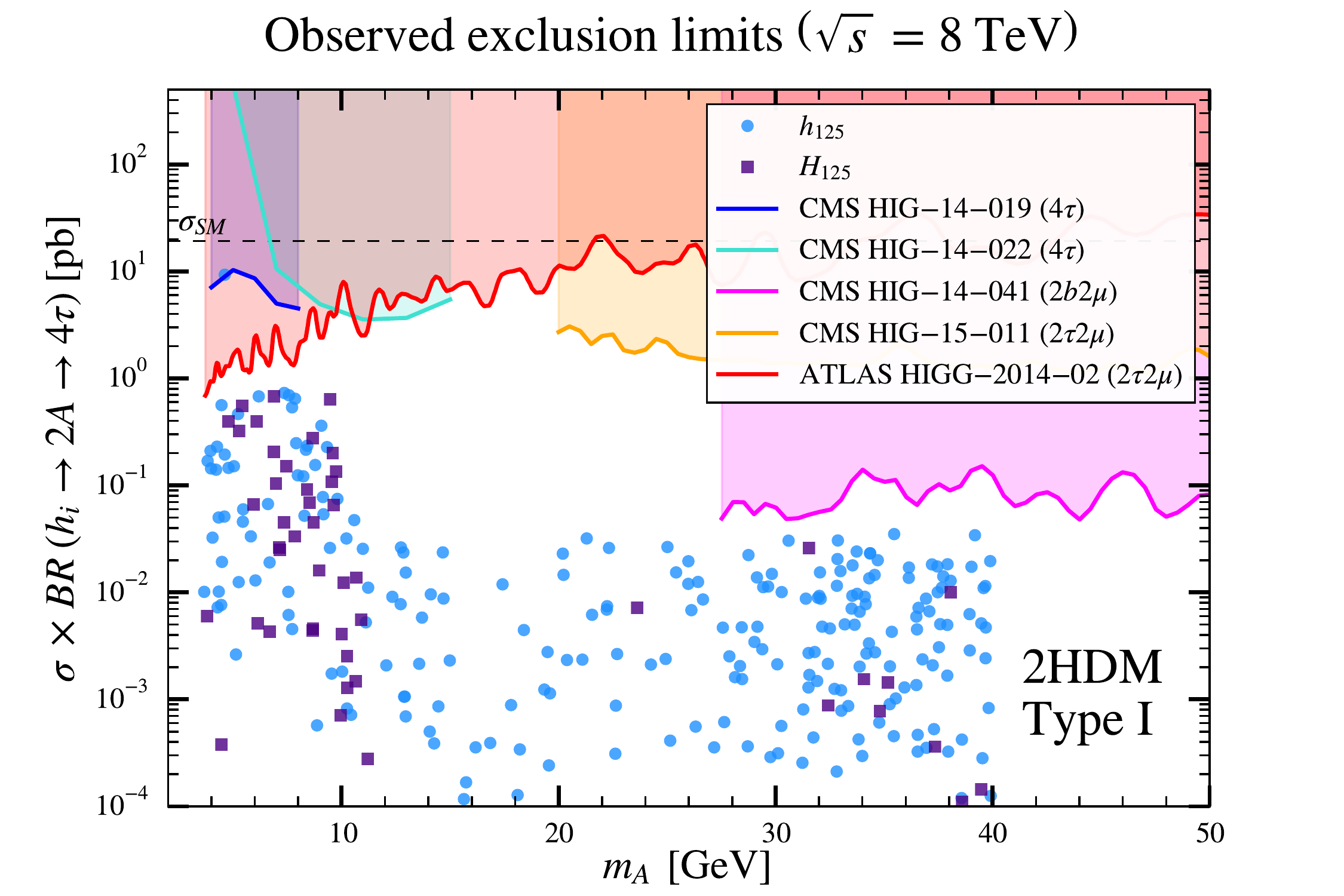}{}\hfill
  \includegraphics[width=0.48\textwidth]{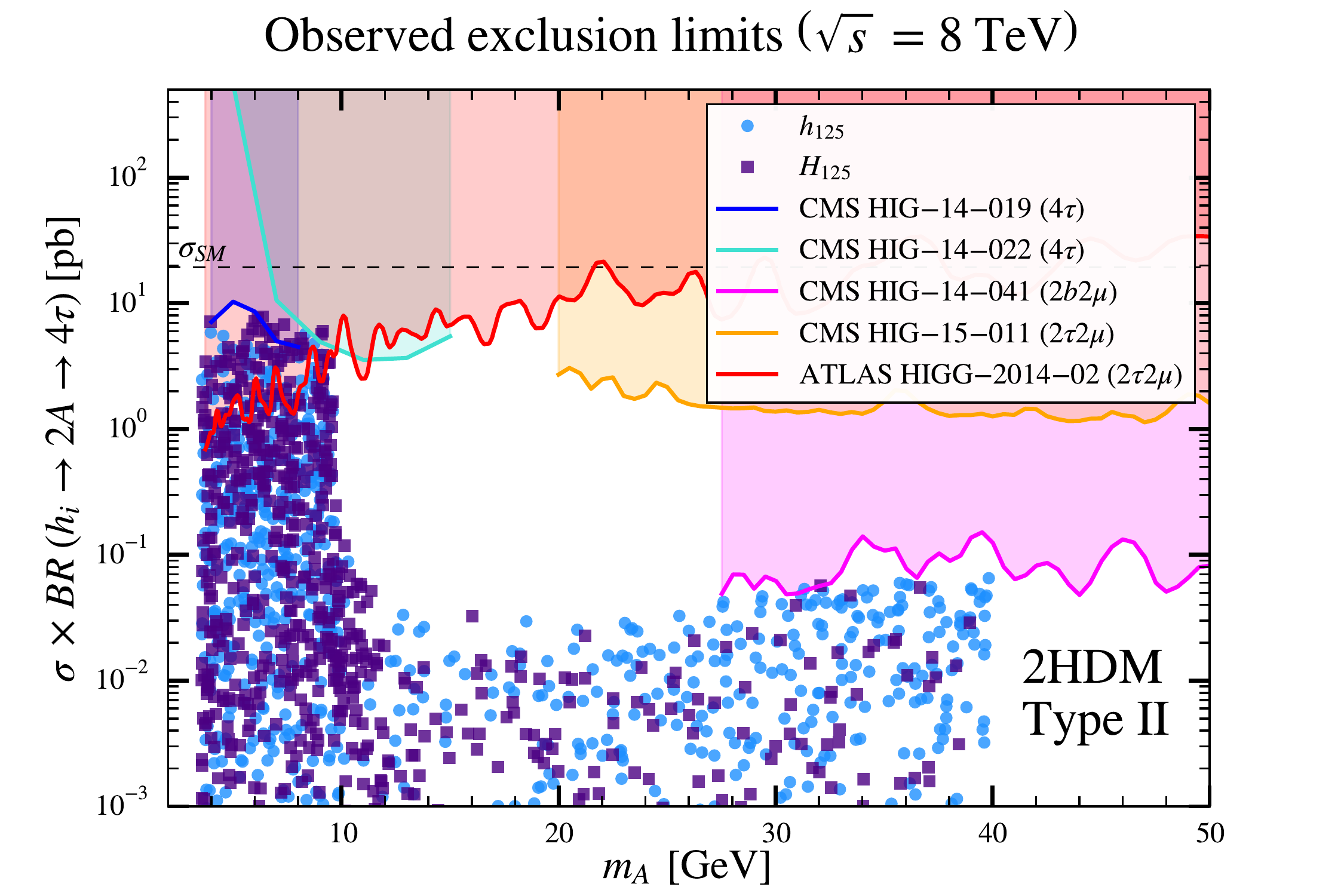}{}l
 \caption{Plots of $\sigma \times BR(h^{\rm SM} \to 2 A \to 4 \tau)$ versus $m_A$ for various Higgs assignments in the 2HDM. 
The left panel refers to the case of type I Yuakwa coupling, while the right plot refers with type II Yukawa couplings.  We refer to~\cite{Aggleton:2016tdd} for details of the scan constraints indicated in the plot.}
 \label{fig:2hdm}
 \end{figure}

Our findings are presented in Fig.~\ref{fig:nmssm} and Fig.~\ref{fig:2hdm} were we show the reach of the most relevant 
8 TeV low mass scalar searches on the NMSSM/nMSSM and 2HDM parameter space respectively, considering the cases where both the lightest and next-to-the lightest CP-even scalar ($h_1$ and $h_2$) can be identified with the SM Higgs boson (note that in the case of the nMSSM, the SM Higgs boson always correspond to $h_2$ in the region with a light $a_1$~\cite{Barducci:2015zna}).

The results show that a combination of the searches described in Sec.~\ref{Sec:Exp} are able to exclude portions of both the NMSSM and 2HDM parameter space, limited to the case of type II Yukawa coupling for the latter, for $A$ bosons with a mass smaller than 10 GeV, leaving however the region with $m_A>10$ GeV almost unconstrained for the 2HDM and NMSSM scenario, while no constraint is imposed in the nMSSM case, due to the dominance of the $a_1\to \chi^0_1 \chi^0_1$ decay channel direct consequence of the almost singlino nature of both $a_1$ and the lightest SUSY particle $\chi^0_1$. A sensitive increase of the experimental sensitivity is expected during the present run of the LHC, which will then further test the existence of of low mass bosons in multi Higgs model.

\section{Conclusions}
\label{Sec:Conc}

In this talk we have assessed  the status of direct searches for a light neutral Higgs boson in popular BSM scenarios with two Higgs doublets, with a mass ranging from 0.25 to 60 GeV, i.e. in a mass region where the recently discovered 125 GeV Higgs boson could decay into a light scalar, giving then rise to 4 SM fermions final state topology.  Specifically we have established that a combination of the aforementioned searches is able to exclude portions of the NMSSM and 2HDM parameter space with 
$m_A<10$ GeV, leaving however the high mass region unconstrained, while no constraints can be imposed for the nMSSM case. 
We expect the analyses that will be performed during the present run of the LHC to greatly improve the reach on the described channels, thus further testing the existence of low mass scalar bosons.

\bibliography{refs}

\end{document}